\documentstyle[12pt]{article}
\begin{document}
\def\l{\lambda}

\begin{flushright}
UT-Komaba 97-5 \\
\end{flushright}
\vskip 0.5in

\begin{center}
{\Large{\bf Solitons in Chern-Simons Theories of Nonrelativistic
CP$^{N-1}$ Models: }}   \\
{\Large{\bf Spin Textures in the Quantum Hall Effect}}
\vskip 0.7in
{\Large Ikuo Ichinose\footnote{e-mail address:
ikuo@hep1.c.u-tokyo.ac.jp}  and Akira Sekiguchi}

\vskip 0.2in
Institute of Physics, University of Tokyo, Komaba, Tokyo, 153 Japan\\
\end{center}
\vskip 0.2in
\begin{center} 
\begin{bf}
Abstract
\vskip 0.5in
\end{bf}
\end{center} 
Topological solitons in CP$^{N-1}$ models coupled with a Chern-Simons gauge field
and a Hopf term are studied both analytically and numerically. 
These models are low-energy effective theories for the quantum Hall effect (QHE)
with internal degrees of freedom, like the QHE in double-layer electron systems.
These solitons correspond to skyrmions and merons which are charged
quasi-excitations in the QHE.
We explicitly show that the CP$^{N-1}$ models describe quite well (pseudo-)spin
textures in the original Chern-Simons theory of bosonized electrons.

\newpage

\section{Introduction}
\setcounter{footnote}{0}
Recently, topological solitons in the Chern-Simons (CS) gauge theories
of nonlinear $\sigma$-models are studied rather intensively\cite{soliton}.
These solitons generally have both fractional spin and statistics.
In the present paper, we shall consider nonrelativistic counterparts of these
models, i.e., nonrelativistic CP$^{N-1}$ (NRCP$^{N-1}$) models coupled 
with CS gauge field.

In the previous paper\cite{IS} (hereafter referred to as I), 
we studied the fractional quantum Hall effect (FQHE)
in double-layer (DL) electron systems in terms of CS gauge theory of bosonized
electrons (CSBE).
There quasi-excitations are topological solitons with fractional spin and statistics,
and we studied them rather intensively.
We also showed that a NRCP$^{1}$ model appears as an effective low-energy
field theory and the above solitons can be reinterpreted as solitons in the 
NRCP$^{1}$ model\cite{IS,Hansson}. 
This NRCP$^{1}$ model couples with a CS gauge field and has a Hopf term.
Therefore solitons have fractional spin and statistics.

In this paper, we shall first study topological solitons in the NRCP$^{1}$ model
for the FQHE in DL electron systems.
It is important to know how these solitons in the above two different CS theories 
are related with each other, not only qualitatively but also quantitatively.
In the CSBE for the DL electron systems, there are two CS gauge fields generally.
In order to obtain the effective theory of nontrivial pseudo-spin textures like merons,
i.e., the NRCP$^1$ model, one CS gauge field is integrated out.
As a result, a Hopf term appears which contributes to spin and statistics
of solitons.
The remaining CS gauge field in the NRCP$^1$ model also contributes
to spin and statistics.
Total spin and statistics of solitons must be consistent with those  obtained by
the argument in terms of the Aharonov-Bohm (AB) effect in the original  CSBE.

Then, we shall consider generalized models, i.e., CS gauge theories of 
NRCP$^3$ models which are effective field theories for spin textures in
the DL QHE with real spin degrees of freedom.
Qualitative and quantitative investigations on solitons are given.

This paper is organized as follows.
In Sec.2, we shall briefly review  the CSBE and the NRCP$^1$ model,
and also properties of solitons in the CSBE.
In Sec.3, solitons in the NRCP$^1$ model are studied and compared with
solitons in the CSBE.
Spin and statistics of solitons are obtained.
By the numerical calculation, we obtain explicit form of them. 
These results show that solitons in the above two CS theories are in good agreement
with each other not only qualitatively but also {\em quantitatively}.
In Sec.4., the NRCP$^3$ models coupled with CS gauge fields are introduced.
Topological solitons are studied both qualitatively and quantitatively.
Section 5 is devoted for conclusion.


\section{CSBE and NRCP$^{N-1}$ model}
\setcounter{equation}{0}

In the paper I, we studied the CSBE for DL FQHE, whose
Lagrangian is given by,
\begin{eqnarray}
{\cal L}&=&{\cal L}_{\psi}+{\cal L}_{\mbox{\small CS}}, \nonumber  \\
{\cal L}_{\psi}&=&-\bar{\psi}_{\uparrow}(\partial_0-ia^+_0-ia^-_0)\psi_{\uparrow}  
-\bar{\psi}_{\downarrow}(\partial_0-ia^+_0+ia^-_0)\psi_{\downarrow}   \nonumber   \\
&&\; -{1 \over 2M}\sum_{\sigma=\uparrow,\downarrow}
|D^{\sigma}_j\psi_{\sigma}|^2-V[\psi_{\sigma}],  
\label{Lpsi}   \\
{\cal L}_{\mbox{\small CS}}&=&{\cal L}_{\mbox{\small CS}}(a^+_{\mu})+
{\cal L}_{\mbox{\small CS}}(a^-_{\mu})   \nonumber   \\
&=&-{i\over 4}\epsilon_{\mu\nu\lambda}\Big({1\over p}a^+_{\mu}
\partial_{\nu}a^+_{\lambda}
+{1\over q}a^-_{\mu}\partial_{\nu}a^-_{\lambda}\Big),
\label{LCS}
\end{eqnarray}
where $\psi_{\sigma}$ ($\sigma=1,2$ or $\uparrow,\downarrow$) is the bosonized 
electron fields in upper and lower layers,
respectively, $M$ is the mass of electrons and $p$ and $q$ are parameters.
Greek indices take $0,1$ and $2$, while roman indices take $1$ and $2$.
$\epsilon_{\mu\nu\lambda}$ is the antisymmetric tensor and the
covariant derivative is defined as 
\begin{equation}
D^{\uparrow\downarrow}_j=\partial_j-ia^+_j\mp ia^-_j+ieA_j,
\end{equation}
where external magnetic field is directed to the $z$-axis and in the symmetric Coulomb gauge
$A_j=-{B \over 2}\epsilon_{jk}x_k$.
$V[\psi_{\sigma}]$ is interaction term between electrons like the Coulomb repulsion
and short-range four-body interaction, e.g.,
\begin{equation}
g_1\Big(|\psi_{\uparrow}|^4+|\psi_{\downarrow}|^4\Big)+g_2|\psi_{\uparrow}
\psi_{\downarrow}|^2, \;\;  \mbox{etc.}
\label{interaction}
\end{equation}
Chern-Simons constraints are obtained by differentiating the Lagrangian
with respect to $a^{\pm}_0$
\begin{eqnarray}
\epsilon_{ij}\partial_ia^+_j&=&2p\bar{\Psi}\Psi  \nonumber   \\
\epsilon_{ij}\partial_ia^-_j&=&2q\bar{\Psi}\sigma_3\Psi  \nonumber   \\
\Psi&=&(\psi_{\uparrow} \;\; \psi_{\downarrow})^t,
\label{CScon}
\end{eqnarray}
where $\epsilon_{ij}=\epsilon_{0ij}$.
As we explained in the paper I , $(p+q)$ must be an odd integer times $\pi$ and $(p-q)$
must be an integer times $\pi$ for the original electrons to be fermionic.

Partition function in the imaginary-time formalism is given as 
\begin{equation}
Z=\int [D\bar{\psi}D\psi Da]\exp \{\int d\tau d^2x ({\cal L}_{\psi}+{\cal L}_{\mbox{\small CS}})\}.
\label{partition}
\end{equation}
The Lagrangian ${\cal L}={\cal L}_{\psi}+{\cal L}_{\mbox{\small CS}}$ is invariant under 
$U(1)\otimes U(1)$ gauge transformation,
\begin{eqnarray}
\Psi &\rightarrow& e^{i\theta_1+i\theta_2\sigma_3}\Psi  \nonumber    \\
a^+_{\mu}  &\rightarrow&  a^+_{\mu}+\partial_{\mu}\theta_1   \label{gaugetrn}  \\
a^-_{\mu}  &\rightarrow&  a^-_{\mu}+\partial_{\mu}\theta_2.  \nonumber
\end{eqnarray}

Ground state for FQHE is given by the following
field configuration,
\begin{eqnarray}
&&\psi_{\uparrow,0}=\sqrt{\bar{\rho}_{\uparrow}}=\sqrt{\bar{\rho}/2},  \nonumber  \\
&&\psi_{\downarrow,0}=\sqrt{\bar{\rho}_{\downarrow}}=\sqrt{\bar{\rho}/2},  \nonumber  \\
&&eB=\epsilon_{ij}\partial_ia^+_{j,0}=2p\bar{\rho}, \; \; \epsilon_{ij}\partial_ia^-_{j,0}=0, 
\;\; a^+_{j,0}=eA_j,  \label{groundcon}   \\
&& \nu={2\pi\bar{\rho} \over eB}={\pi \over p},
\label{ground}
\end{eqnarray}
where $\bar{\rho}$ is the average electron density.
It is easily verified that the above static and uniform configuration is a solution 
to the field equations
if and only if the filling factor $\nu$ has specific value given by (\ref{ground}).

We assume the following specific form of the potential between bosonized
electrons;
\begin{equation}
V[\psi_{\sigma}]={p\over M}\Big(\bar{\Psi}\Psi\Big)^2+
{q\over M}\Big(\bar{\Psi}\sigma_3\Psi\Big)^2.
\label{pot}
\end{equation}
It is known that the above short-range repulsions represent the Pauli exclusion 
principle in the bosonization of fermion in the CS method\cite{EHI}.

From (\ref{pot}) and the Bogomol'nyi decomposition,
the Hamiltonian is given by
\begin{equation}
{\cal H}={1\over 2M}\sum_{\sigma}|(D^{\sigma}_1-iD^{\sigma}_2)\psi_{\sigma}|^2
+{\omega_c \over 2}\Big(\bar{\Psi}\Psi\Big),
\label{Hamiltonian}
\end{equation}
where $\omega_c=eB/M$.
Therefore, the lowest-energy configurations satisfy the following ``self-dual"
equations,
\begin{equation}
(D^{\uparrow}_1-iD^{\uparrow}_2)\psi_{\uparrow}=0, \;
(D^{\downarrow}_1-iD^{\downarrow}_2)\psi_{\downarrow}=0.
\label{self}
\end{equation}
 
Topological solitons are solutions to the self-dual equations (\ref{self})\cite{SDCST}.
We consider only spherical symmetric configurations and therefore they are
parameterized as follows,
\begin{eqnarray}
\psi_{\uparrow}&=&\sqrt{\bar{\rho}}\exp \Big(w_1(r)+in_1\theta (x)\Big),  \nonumber  \\
\psi_{\downarrow}&=&\sqrt{\bar{\rho}}\exp \Big(w_2(r)+in_2\theta (x)\Big),
\label{para1}
\end{eqnarray}
where $r=|\vec{x}|$, $\theta(x)$ is the azimuthal function 
$\theta(x)=\arctan \Big({x_2\over x_1}\Big)$,
and $n_1$ and $n_2$ are integers which label topological solitons.
It is rather straightforward to show that the self-dual equations (\ref{self})
give the following Toda-type equations of $w_i(r) \; (i=1,2)$,
\begin{equation}
 {d^2 w_i\over d \l^2}+{1\over \l}{d w_i\over d \l}
+n_i{\delta (\l) \over \l}+2-2\sum_j K_{ij}e^{2w_j}=0,
\label{fieldeq1}
\end{equation}
\begin{equation}
 \l=\sqrt{p\bar{\rho}}\cdot r={r \over \sqrt{2}l_0}, \; \; K={1\over p} \left(
                                         \begin{array}{cc}
					 p+q & p-q  \\
					 p-q & p+q
					 \end{array}
					 \right),  \nonumber 
\end{equation}
where $l_0$ is the magnetic length $l_0=1/\sqrt{eB}$.

It is shown\cite{IS} that the integers $n_i$ must be nonpositive,
$n_i=0,-1,-2,...$, for single-valuedness and regularity of solution at the center of soliton.
On the other hand, boundary condition at the infinity is given as 
\begin{equation}
\lim_{\l\rightarrow \infty}e^{w_i(\l)}={1\over \sqrt{2}}.
\label{bc}
\end{equation}

In the paper I, we showed that charges, $Q$ and $\bar{Q}$,
and spin of solitons, $S(\mbox{soliton})$, are given as 
\begin{eqnarray}
&& Q\equiv\int d^2x\Big( \bar{\Psi}\Psi -\bar{\Psi}_0\Psi_0\Big)   
={\pi \over p}\Big({n_1+n_2 \over 2}\Big),   \nonumber  \\
&& \bar{Q}\equiv\int d^2x\Big( \bar{\Psi}\sigma_3\Psi -\bar{\Psi}_0\sigma_3\Psi_0\Big)
= {\pi \over q}\Big({n_1-n_2 \over 2}\Big),       \nonumber  \\
&& S(\mbox{soliton})=S_++S_-,  \nonumber   \\
&& S_+={\pi \over 2p}\Big({n_1+n_2\over 2}\Big)^2,  \; \; 
 S_-={\pi \over 2q}\Big({n_1-n_2\over 2}\Big)^2.
\label{QJ}
\end{eqnarray}
On the other hand, statistical parameter is given as follows by the argument of AB effect,
\begin{eqnarray}
&& \alpha(\mbox{soliton})=\alpha_++\alpha_-,  \nonumber  \\
&& \alpha_+=-pQ^2,  \nonumber   \\
&& \alpha_-=-q\bar{Q}^2.
\label{statistics}
\end{eqnarray}
Therefore, the spin-statistics relation $S=-{1 \over 2\pi}\alpha$ is satisfied.

In the paper I, we solved the self-dual equations (\ref{self}) for various values of
$p,q,n_1$ and $n_2$ by the numerical calculation.
Some of them are given in the paper I and Figs.1 and 2 of the present paper.
These solutions will be compared with their counterparts in the NRCP$^1$ model later on.
To this end, normalized amplitudes of $\psi_{\sigma}$'s are defined as 
\begin{equation}
\hat{\psi}_{\sigma}\equiv {|\psi_{\sigma}| \over \sqrt{|\psi_{\uparrow}|^2+
|\psi_{\downarrow}|^2}}.
\label{normal}
\end{equation}

In order to obtain an effective field theory for topological excitations from the CSBE,
we first parameterize the bosonized electrons as follows;
\begin{eqnarray}
\psi_{\sigma}&=&J^{1/2}_0\phi \; z_{\sigma},  \nonumber  \\
 \phi&=&e^{i\tilde{\theta}}\phi_{v}\;\in\;  U(1), \;\; z=(z_1 \;\; z_2)^t\; \in \; \mbox{CP}^1,
\label{para}
\end{eqnarray}
where $e^{i\tilde{\theta}}$ is a regular part and $\phi_v$ is a topologically nontrivial part, 
which represents vortex degrees of freedom.
CP$^1$ variable $z_{\sigma}$ represents pseudo-spin degrees of freedom and satisfies
CP$^1$ condition $\sum |z_{\sigma}|^2=1$, and $\psi_{\sigma}$ is
invariant under $z_{\sigma}\rightarrow e^{i\alpha}z_{\sigma}$ and 
$\phi \rightarrow e^{-i\alpha}\phi$.
After substituting (\ref{para}) into (\ref{Lpsi}), we perfom duality transformation
in the partition function (\ref{partition})\footnote{Here we neglect the potential
term $V[\psi_{\sigma}]$ in (\ref{Lpsi}).
Interaction terms in the NRCP$^1$ model will be discussed in Sec.3.}\cite{Lee}. 
Details of the derivation can be seen in the paper I.
Final result is given by the following Lagrangian;
\begin{eqnarray}
{\cal L}_E&=&{\cal L}_z+{\cal L}_{\mbox{\small CS}}(a^-_{\mu})
-ib_i\cdot \hat{J}_i,    \nonumber  \\
{\cal L}_z&=&-{J_0\over 2M}\Big[\overline{D_jz}\cdot D_jz+
(\bar{z}\cdot D_jz)^2\Big],   \label{effeL}  \\
D_{\mu}z&=&(\partial_{\mu}-ia^-_{\mu}\sigma_3)z, \nonumber
\end{eqnarray}
where $\hat{J}_{\mu}=J^v_{\mu}+J^S_{\mu}$ is the sum of topological currents
of vortex and pseudo-spin texture and they are explicitly given as 
\begin{eqnarray}
J^v_{\mu}&=&{1\over 2\pi}\epsilon_{\mu\nu\lambda}\partial_{\nu}\Big(\bar{\phi}_v
{\partial_{\lambda} \over i}\phi_v\Big),   \label{Jv}  \\
J^S_{\mu}&=&{1\over 2\pi}\epsilon_{\mu\nu\lambda}\partial_{\nu}\Big(\bar{z}
{D_{\lambda} \over i}z\Big),   \label{JS} 
\end{eqnarray}
and the vector field $b_i$ is related with the total electron density $J_0$ in (\ref{para})
as
\begin{equation}
J_0={1 \over 2\pi} \epsilon_{ij}\partial_ib_j.
\label{bj}
\end{equation}
Furthermore, we are interested in only the states in the lowest Landau level (LLL).
The LLL condition imposes\footnote{More precisely as we showed in the paper I, 
there appear additional higher-derivative terms like  $\epsilon_{ij}\partial_iJ^S_j$
in the LLL conditon.
However {\em total charge} of soliton is determined solely by its topological charge (see the
discussion in the paper I).}\cite{IS,LLL}
\begin{equation}
J^v_0+J^S_0-\nu^{-1}J_0+{e\over 2\pi}\epsilon_{ij}\partial_iA_j=0.
\label{LLL}
\end{equation}
Therefore, the topological charges determine the electric charge of
solitons.
From (\ref{bj}) and (\ref{LLL}), the last term of ${\cal L}_E$ in (\ref{effeL}) 
containes the Hopf term, which gives spin and statistics to solitons\cite{WZ}.
It is not so difficult to show that these are $S_+$ and $\alpha_+$
in (\ref{QJ}) and (\ref{statistics}), respectively.

We have discussed the case of two-component internal degrees of freedom so far.
It is rather straightforward to extend this formalism.
For example, when we consider DL electron systems with spin degrees of freedom,
internal space has a (approximate) SU(4) symmetry, and nontrivial spin-pseudo-spin
textures are described by a NRCP$^3$ model.


\section{Solitons in the NRCP$^1$ model}
\setcounter{equation}{0}

In Sec.2, we briefly reviewed the relationship between the CSBE ((\ref{Lpsi}) and (\ref{LCS}))
and the NRCP$^1$ model ((\ref{effeL})).
In this section we shall study topological solitons in the NRCP$^1$ model
rather in detail.

The last term of ${\cal L}_E$ in (\ref{effeL}) is rewritten as follows,
\begin{eqnarray}
\int d^2x b_i\cdot \hat{J}_i&=&{1\over 2\pi}\int d^2x \epsilon_{ij}\partial_0 b_i
\Big\{\bar{\phi}_v{\partial_j \over i}\phi_v+\bar{z}\cdot {D_j\over i}z\Big\}  \nonumber  \\
  && +{1\over 2\pi}\int d^2x\epsilon_{ij}b_i\Big\{\partial_j(\bar{\phi}_v{\partial_0\over i}
  \phi_v)+\partial_j(\bar{z}\cdot {D_0 \over i}z)\Big\}.
\label{intbJ}
\end{eqnarray}  
As we consider static configurations in the following discussion, the first two terms on 
the right-hand side of (\ref{intbJ}) are ignored.
From the LLL condition (\ref{LLL}),
\begin{equation}
J_0={1\over 2\pi}\epsilon_{ij} \partial_ib_j=\bar{\rho}+{\pi \over p} (J^v_0+J^S_0).
\label{LLL2}
\end{equation}
By substituting (\ref{LLL2}) into (\ref{intbJ}), we obtain
\begin{equation}
{\cal L}_E=J_0(\bar{\phi}_vi\partial_0\phi_v)+J_0(\bar{z}\cdot iD_0z)
-{J_0\over 2M}\{|D_jz|^2+(\bar{z}\cdot D_jz)^2\}-{1\over 4q}\epsilon_{\mu\nu\lambda}
a^-_{\mu}\partial_{\nu}a^-_{\lambda}.
\label{effeL2}
\end{equation}
The Lagrangian ${\cal L}_E$ is invariant under local $U(1)\otimes U(1)$
transformation,
\begin{eqnarray}
z & \rightarrow & e^{i\phi_1+i\phi_2\sigma_3}z,  \nonumber  \\
a^-_{\mu} & \rightarrow & a^-_{\mu}+\partial_{\mu}\phi_2. 
\label{CPgauge}
\end{eqnarray}
Field eqution of $a^-_0$ gives the CS constraint
\begin{equation}
{1\over 2q}\epsilon_{ij}\partial_ia^-_j=J_0(\bar{z}\cdot \sigma_3z), 
\label{CScon2}
\end{equation}
which corresponds to (\ref{CScon}) in the CSBE.

In the following discussion, we shall put $J_0=\bar{\rho}$ in the Lagrangian
(\ref{effeL2}).
This approximation corresponds to neglecting higher-derivative terms 
of the field $z_j$.
Then the Hamiltonian density is obtained as
\begin{eqnarray}
{\cal H}^0_E&=& {\bar{\rho} \over 2M}\{|D_jz|^2+(\bar{z}\cdot D_jz)^2\}  \nonumber   \\
  &=&{\bar{\rho} \over 2M}|\bigtriangledown_jz|^2, 
\label{H2}
\end{eqnarray}
where we have defined $\bigtriangledown_{\mu}z=D_{\mu}z-(\bar{z}\cdot D_{\mu}z)z$.

In order to make the Hamiltonian Bogomol'nyi type, we add the following
potential term to the Hamiltonian ({\ref{H2}),
\begin{equation}
U[z]={q \over M}\bar{\rho}^2(\bar{z}\cdot \sigma_3 z)^2.
\label{pot2}
\end{equation}
It is obvious that this term corresponds to the second term of the potential 
$V[\psi_{\sigma}]$ in (\ref{pot}).
One may conceive that potential corresponding to the first term of $V[\psi_{\sigma}]$ 
should be also added.
However from (\ref{LLL2}),
\begin{eqnarray}
\Big(\bar{\Psi}\Psi\Big)^2&=&J_0^2  \nonumber  \\
&=& \Big[ \bar{\rho}^2+{2\bar{\rho}\pi \over p}(J^v_0+J^S_0)+\Big({\pi \over p}\Big)^2
(J^v_0+J^S_0)^2\Big].
\label{Psi2}
\end{eqnarray}
In Eq.(\ref{Psi2}), the term proportional to $J^S_0$ gives just the topological number 
and it does not contribute to the field equations, and the term $(J^S_0)^2$ is 
higher derivative of the field $z_j$.

Using the identity,
\begin{equation}
|\bigtriangledown_jz|^2=|(\bigtriangledown_1-i\bigtriangledown_2)z|^2
-\{2\pi J^S_0+i\epsilon_{kl}\bar{z}\cdot \partial_k\partial_l z+\epsilon_{ij}
\partial_ia^-_j (\bar{z}\cdot \sigma_3z)\},
\label{Bid}
\end{equation}
the total Hamiltonian is given as 
\begin{eqnarray}
{\cal H}_E&=& {\cal H}^0_E+U[z]    \nonumber  \\
                &=& {\bar{\rho} \over 2M}|(\bigtriangledown_1-i\bigtriangledown_2)z|^2
		   -{\pi\bar{\rho} \over M}J^S_0.
\label{H3}
\end{eqnarray}
We have omitted the possible $\delta$-function type potential,
which may arise from the singularity of $z$ at the origin, i.e.,
$ \epsilon_{kl}\bar{z}(x)\cdot \partial_k\partial_l z(x) \propto \delta (x) $.
This is legitimate, for solitons in the original CSBE with both nonvanishing winding
numbers $n_1$ and $n_2$ have vanishing amplitude at the origin, $J_0(x=0)=0$,
and therefore
\begin{equation}
J_0(x)\cdot \bar{z}(x)\cdot \epsilon_{kl}\partial_k\partial_l z(x)=0, \nonumber
\end{equation}
for arbitrary $x$.

From (\ref{H3}), the lowest-energy configurations satisfy the following 
self-dual equations;
\begin{equation}
(\bigtriangledown_1-i\bigtriangledown_2)z=0.
\label{self2}
\end{equation}
Since $z$ is complex field of two component, the equation (\ref{self2})
gives four equations.
However, it is easily shown that two of them are redundant by the constraint
$\bar{z}\cdot z=1$.

``Gound state" of the self-dual equation (\ref{self2}) is given as 
\begin{equation}
z_G={1\over \sqrt{2}}\left(
                                  \begin{array}{c}
				  1   \\
				  1
				  \end{array}
				  \right),
\label{zG}
\end{equation}				  
up to the $U(1)\otimes U(1)$ symmetry (\ref{CPgauge}).
We parameterize configurations as follows;
\begin{equation}
z=\left(
         \begin{array}{c}
        X(r)e^{in_1\theta}  \\
	Y(r)e^{in_2\theta}
        \end{array}
	\right),
\label{confz}
\end{equation}
where $(r,\theta)$ is the polar coordinates as before.
We assume that $X(r)$ and $Y(r)$ are real functions which depend only on $r$,
and $n_1$ and $n_2$ are integers which label solitons, as in the CSBE. 
At sufficiently large distance, 
\begin{equation}
X(r) \rightarrow {1 \over \sqrt{2}}, \; Y(r) \rightarrow {1 \over \sqrt{2}}, \; \mbox{as} \;
r \rightarrow +\infty.
\label{Bcon}
\end{equation}
From (\ref{confz}) and (\ref{Bcon}), the CS gauge field must behave as 
\begin{equation}
a^-_j \rightarrow \partial_j \theta \cdot {1 \over 2}(n_1-n_2), \; \mbox{as} \;
r \rightarrow +\infty,
\label{aBcon}
\end{equation}
for the energy of configuration, which is obtained from (\ref{H3}), to be finite.
From (\ref{aBcon}), it is obvious that the CS flux of the $(n_1,n_2)$ soliton
is given as 
\begin{equation}
\int d^2x \epsilon_{ij}\partial_ia^-_j=2\pi \Big({n_1-n_2 \over 2}\Big).
\label{CSflux}
\end{equation}
Therefore from the CS constraint (\ref{CScon2}), the ``axial" charge of the 
CP$^1$ soliton is 
\begin{eqnarray}
\bar{Q}_z&\equiv& \bar{\rho}\int d^2x(\bar{z}\cdot \sigma_3z)  \nonumber  \\
&=&{\pi \over q} \Big({n_1-n_2 \over 2}\Big)   \nonumber   \\
&=& \bar{Q}, 
\label{Qz}
\end{eqnarray}
where the last equality comes from (\ref{QJ}).
Then the argument by the AB effect indicates that the CS gauge field $a^-_{\mu}$
gives statistical parameter to the soliton as 
\begin{equation}
-q\bar{Q}_z=\alpha_-.
\label{stapara2}
\end{equation}
We already mentioned that the Hopf term $b_i\cdot J^S_i$ in (\ref{effeL})
gives $\alpha_+$, and therefore the soliton in the NRCP$^1$ has the 
same statistical parameter with that of corresponding soliton in the CSBE,
as it should be.

Substituting (\ref{confz}) into (\ref{self2}), we obtain the following two
independent field equations;
\begin{equation}
\partial_k\ln (X^{-2}-1)^{{1\over 2}}=(n_1-n_2)\epsilon_{kl}\partial_l\theta
-2\epsilon_{kl}a^-_l, \; \; k=1,2.
\label{fieldeq2}
\end{equation}
Differntiating Eq.(\ref{fieldeq2}) by $\partial_k$ and using the constraint
(\ref{CScon2}),
\begin{equation}
\sum_k\partial_k^2   \ln (X^{-2}-1)^{{1 \over 2}}+4q\bar{\rho}(2X^2-1)=2\pi (n_1-n_2)
\delta^{(2)}(x).
\label{fieldeq3}
\end{equation}
It is useful to introduce the function $u={1\over 2}\ln (X^{-2}-1)$ and
$\lambda=\sqrt{p\bar{\rho}}r={r \over \sqrt{2}l_0}$, 
and Eq.(\ref{fieldeq3}) gives\footnote{It should be remarked that the NRCP$^1$
model {\em does not} contain the parameter $p$ and its solitons are independent of
the value of $p$ in the original CSBE.
Therefore using $\lambda$ is merely for convenience sake,
and $\lambda$ measures the distance in the unit of the magnetic length $l_0$.}
\begin{equation}
\Big({1\over \lambda}{d\over d\lambda}\lambda{d\over d\lambda}\Big)
u(\lambda)+4\Big({p\over q}\Big){1-e^{2u} \over 1+e^{2u}}=
(n_1-n_2){\delta(\lambda) \over \lambda}.
\label{fieldeq4}
\end{equation}
In order to remove the $\delta$-function singularity on the right-hand
side of (\ref{fieldeq4}), we furthermore rewrite it in terms of $v(\lambda)
=u(\lambda)-\ln \lambda^{(n_1-n_2)}$.
Field equation of  $v(\lambda)$ is then given as
\begin{equation}
\Big({1\over \lambda}{d\over d\lambda}\lambda{d\over d\lambda}\Big)
v(\lambda)+4\Big({p\over q}\Big)\Bigg\{{1-\lambda^{2(n_1-n_2)}e^{2v(\lambda)}
\over 1+\lambda^{2(n_1-n_2)}e^{2v(\lambda)}}\Bigg\}=0.
\label{fieldeq5}
\end{equation}
In terms of $v(\lambda)$,
\begin{equation}
X^2={1 \over 1+\lambda^{2(n_1-n_2)}e^{2v(\lambda)}}, \;\;
Y^2={1 \over 1+\lambda^{-2(n_1-n_2)}e^{-2v(\lambda)}}.
\label{XY}
\end{equation}
From (\ref{XY}), it is obvious that at the spatial infinity,
\begin{equation}
v(\lambda) \sim -(n_1-n_2) \ln \lambda, \;\; \lambda \rightarrow \mbox{large}.
\label{BCv}
\end{equation}
On the other hand at the origin, $v(\lambda)$ behaves as 
\begin{equation}
v(\lambda) \sim \beta (n_1,n_2)-{\lambda^2 \over 2},
\label{BCv2}
\end{equation}
where the parameter $\beta (n_1,n_2)$ depends on the type of soliton and it is
determined by the requirement that the solution satisfies the boundary condition at the spatial
infinity (\ref{BCv}) and (\ref{Bcon}).

A comment on the solutions to field equation (\ref{fieldeq5}) is in order.
In the usual case, amplitude of field which has nontrivial winding number must vanish
at the origin by the requirement of the uniqueness and the single-valuedness.
Therefore one may conceive that topological solitons which have 
both nonvanishing winding numbers $n_1$ and $n_2$ cannot exist in the present model,
for vanishing of both amplitudes, $X(r)$ and $Y(r)$, contradicts the CP$^1$ condition.
However as we showed in the paper I and Sec.2 of the present paper, 
in the original CSBE the sum of electron densities in the upper and lower layers 
of the above type solutions vanishes 
at the origin, i.e., $J_0(r=0)=0$ for $n_1\neq 0$ and $n_2 \neq 0$, and therefore
the solution is meaningful even if the CP$^1$ part violates the uniqueness condition.
Then we do not require the uniqueness condition for the solutions  in the NRCP$^1$
model and consider, for example, $(-2, -1)$ soliton, in which at least one of the components
of the field $z$ does not vanish at the origin.
It should be compared with the corresponding $(-2, -1)$ solution in the CSBE
after {\em normalization}. 

Let us turn to the results of the numerical calculation.
They are given in Figs.1 and 2.
For Halperin's $(m,m,n)$ state, the parameters $p$ and $q$ are given by
$p={\pi \over 2}(m+n)$ and $q={\pi \over 2}(m-n)$, as before.
Corresponding configurations of $z$ field obtained from the solutions in the CSBE 
by normalization (\ref{normal}) are also shown in the figures.
Quantitative agreement of them is obvious.
Therefore, we can conclude that the NRCP$^1$ model describes quite well
pseudo-spin textures in the original CSBE.

Especially, solitons in the CSBE depend on the parameters $p$ and $q$,
whereas those in the NRCP$^1$ model depend on only $q$.
That is, solitons in the CSBE's with same $q$ but with different $p$ correspond
to the same soliton in the NRCP$^1$ model.\footnote{Please recall that value
of $p$ determines the filling factor as (\ref{ground}).
Therefore, this result means that a series of solitons at different filling factors 
in the CSBE have the same form after normalization.}
For example, $(-1,0)$ meron in the $p={5\pi \over 2}$ and $q={\pi \over 2}$ CSBE
and that in the $p={9\pi \over 2}$ and $q={\pi \over 2}$ CSBE both
correspond to $(-1,0)$ meron in the $q={\pi \over 2}$ NRCP$^1$.
We numerically examined this case and found that after normalization
the above two soliton solutions have almost the same form and it coincides
with the $(-1,0)$ meron in the NRCP$^1$ model (see Fig.2).
In this sense, the NRCP$^{N-1}$ models are more universal than the CSBE.


\section{Solitons in the NRCP$^3$ model}
\setcounter{equation}{0}

In the previous section, we studied topological solitons in the NRCP$^1$ model
both qualitatively and numerically.
It was shown that they are in good agreement with those corresponding
to them in the CSBE.
In this section, we shall consider a generalization of the previous model, i.e., 
the NRCP$^3$ model coupled with a pair of CS gauge fields.
This model can be regarded as an effective low-energy theory which describes
nontrivial configurations in the internal space of the DL QHE with real spin
degrees of freedom. 
It is rather straightforward to introduce bosonized electrons of pseudo-spin
suffix $\sigma=\uparrow, \downarrow$ and spin suffix $\eta=\uparrow, \downarrow$,
$\psi_{\sigma,\eta}$.
Then CP$^3$ field $Z$ is introduced as 
\begin{equation}
Z=(z_{\uparrow\uparrow} \; z_{\uparrow\downarrow} \; z_{\downarrow\uparrow} \;
z_{\downarrow\downarrow})^t \;\; \in \;\; \mbox{CP$^3$},
\label{Z}
\end{equation}
\begin{equation}
\psi_{\sigma,\eta}=J^{1/2}_0\phi \; z_{\sigma,\eta} \; .
\label{para2}
\end{equation}

We study NRCP$^3$ model which is defined by the following 
Lagrangian,\footnote{In the present case, we consider only the total spin-singlet and 
pseudo-spin-singlet quantum Hall state.
General cases will be discussed in a future publication.}
\begin{equation}
{\cal L}^{CP^3}_E=J_0(\bar{Z}\cdot iD_0Z)
-{J_0\over 2M}\{|D_jZ|^2+(\bar{Z}\cdot D_jZ)^2\}-\sum_{l=1,2}
{1\over 4q^{(l)}}\epsilon_{\mu\nu\lambda}
a^{(l)}_{\mu}\partial_{\nu}a^{(l)}_{\lambda},
\label{effeL3}
\end{equation}
where the coefficients of the CS terms, $q^{(l)}$ ($l=1,2$), are parameters
which determine the numbers of CS fluxes attaching to the bosonized electrons 
$\psi_{\sigma, \eta}$ and their values
are determined by Coulomb interactions between electrons.\footnote{In 
Ref.\cite{NA}, it is shown by numerical calculation that generalized
Halperin state $\Psi_{lmn}$ is a good variational wave function for the ground state
of DL QHS with real spin degrees of freedom.
The parameters $q^{(l)}$ $(l=1,2)$ in the NRCP$^3$ CS theory 
are related with $(l,m,n)$ as $q^{(1)} \propto (l-n)$ and $q^{(2)} \propto (l-m)$.}
The covariant derivative in (\ref{effeL3}) is defined as 
\begin{equation}
D_{\mu}Z\equiv (\partial_{\mu}-ia^{(l)}_{\mu}\Gamma^{(l)})Z, 
\label{covZ}
\end{equation}
where $(4\times 4)$ matrices $\Gamma^{(l)}$ are given by
\begin{equation} 
\Gamma^{(1)}=\left(
                       \begin{array}{cc}
		       I & 0  \\
		       0 & -I 
		       \end{array}
		       \right),   \;\;
\Gamma^{(2)}=\left(
                       \begin{array}{cc}
		       \sigma_3 & 0  \\
		       0 & \sigma_3
		       \end{array}
		       \right).
\label{Gamma}		       
\end{equation}		       
We add the following repulsive interactions as required by the CS bosonization method,
\begin{equation}
V[Z]=\sum {q^{(l)} \over M}\bar{\rho}^2\Big(\bar{Z}\cdot \Gamma^{(l)}Z
\Big)^2.
\label{VZ}
\end{equation}
Total Hamiltonian is given by
\begin{equation}
{\cal H}^{CP^3}_E={\bar{\rho} \over 2M}|(\bigtriangledown_1-i\bigtriangledown_2)Z|^2
		   -{\bar{\rho} \over 2M}J^S_0.
\label{HCP3}
\end{equation}
Then lowest-energy configurations satisfy the self-dual eqautions,
\begin{equation}
(\bigtriangledown_1-i\bigtriangledown_2)Z=0,
\label{self3}
\end{equation}
where $\bigtriangledown_{\mu}Z=D_{\mu}Z-(\bar{Z}\cdot D_{\mu}Z)Z$.
The system is invariant under the following gauge transformations,
\begin{eqnarray}
Z &\rightarrow & e^{i\alpha_1\Gamma^{(1)}+ i\alpha_2\Gamma^{(2)}}Z, \nonumber  \\
a^{(l)}_{\mu} &\rightarrow & a^{(l)}_{\mu}+\partial_{\mu}\alpha_l, \; l=1,2  \nonumber  \\
&\mbox{and also} &   \nonumber \\
Z &\rightarrow & e^{i\omega}Z.
\label{gaugetr}
\end{eqnarray}
There is a topological current which is invariant under (\ref{gaugetr}),
\begin{equation}
j^S_{\mu}={1 \over 2\pi}\epsilon^{\mu\nu\lambda}\partial_{\nu}
\Big(\bar{Z}{D_{\lambda} \over i}Z\Big).
\label{topcur2}
\end{equation}

As in the previous case, we shall seek topological solitons of the spherical
symmetry,
\begin{equation}
Z=\left(
     \begin{array}{c}
     F(r)e^{in_1\theta}  \\
     G(r)e^{in_2\theta}  \\
     H(r)e^{in_3\theta}   \\
     R(r)e^{in_4\theta}  
   \end{array}
   \right) \; .
\label{paraZ}
\end{equation}
For example, configuration which corresponds to a skyrmion of real spin in the upper layer    
is described by 
\begin{equation}
n_1=-2, \; n_2=n_3=n_4=0.
\label{ni}
\end{equation}
It is useful to parametrize $F(r), G(r), H(r)$ and $R(r)$ as
\begin{eqnarray}
F^2 &=& {e^{2(u_1+u_2)} \over 1+e^{2(u_1+u_2)}+e^{2u_2}+e^{2u_3}}, \nonumber  \\
G^2 &=& {e^{2u_2} \over 1+e^{2(u_1+u_2)}+e^{2u_2}+e^{2u_3}},  \nonumber  \\
H^2 &=& {e^{2u_3}  \over 1+e^{2(u_1+u_2)}+e^{2u_2}+e^{2u_3}},  \nonumber   \\
R^2 &=& {1 \over 1+e^{2(u_1+u_2)}+e^{2u_2}+e^{2u_3}}.
\label{FGHR}
\end{eqnarray}
Substituing (\ref{paraZ}) and (\ref{FGHR}) into the self-dual equation (\ref{self3}),
\begin{eqnarray}
{d^2u_1 \over dr^2}+{1 \over r}{d u_1 \over dr}
+(n_1-n_2){\delta (r) \over r}-4q^{(2)} \cdot
(\bar{Z}\Gamma^{(2)}Z) &=&0,  \nonumber   \\
{d^2u_2 \over dr^2}+{1 \over r}{d u_2 \over dr}
+(n_2-n_4){\delta (r) \over r}-4q^{(1)} \cdot
(\bar{Z}\Gamma^{(1)}Z) &=&0,  \nonumber   \\
 {d^2u_3 \over dr^2}+{1 \over r}{d u_3 \over dr}
+(n_3-n_4){\delta (r) \over r}-4q^{(2)} \cdot
(\bar{Z}\Gamma^{(2)}Z) &=&0,
\label{equi}
\end{eqnarray}
where
\begin{eqnarray}
\bar{Z}\Gamma^{(1)}Z &=& F^2+G^2-H^2-R^2,  \nonumber  \\
\bar{Z}\Gamma^{(2)}Z &=& F^2-G^2+H^2-R^2, \nonumber
\end{eqnarray}
and we have set $\bar{\rho}=1$ for simplicity.

From (\ref{equi}), it is not difficult to get behaviours of $u_i \; (i=1,2,3)$ for 
$r \sim 0$.
For example for soliton $(\{n_i\})=(-1,0,0,0)$
\begin{eqnarray}
u_1 &\sim &\ln r+a_1+q^{(2)} \cdot
\Big\{{-e^{2a_2}+e^{2a_3}-1 \over 1+e^{2a_2}+e^{2a_3}}\Big\}r^2,  \nonumber  \\
u_2 &\sim & a_2+q^{(1)} \cdot
\Big\{{e^{2a_2}-e^{2a_3}-1 \over 1+e^{2a_2}+e^{2a_3}}\Big\}r^2,  \nonumber  \\
 u_3 &\sim & a_3+q^{(2)} \cdot
\Big\{{-e^{2a_2}+e^{2a_3}-1 \over 1+e^{2a_2}+e^{2a_3}}\Big\}r^2,  
\label{asyui}
\end{eqnarray}
where $a_i \; (i=1,2,3)$ are parameters which are determined by the boundary condition
at the spatial infinity.
There are three ``interesting" ground states in the present system;
\begin{equation}
Z_G={1 \over \sqrt{2}}\left(
     \begin{array}{c}
         1 \\
	 0 \\
	 0  \\
	 1
	 \end{array}
	 \right),
	 \;\;
{1 \over \sqrt{2}}\left(
     \begin{array}{c}
         0 \\
	 1 \\
	 1  \\
	 0
	 \end{array}
	 \right),
	 \;\;
{1 \over \sqrt{4}}\left(
     \begin{array}{c}
         1 \\
	 1 \\
	 1  \\
	 1
	 \end{array}
	 \right).
\label{Zground}
\end{equation}
The first one corrresponds to the configuration in which the upper layer is filled with
electrons with up spin and  the lower layer is filled with those with down spin, etc.
General ``ground state" is given by an arbitrary linear combination of the first two
states in (\ref{Zground}).
The parameters $a_i$ in (\ref{asyui}) are determined by the requirement
that at the spatial infinity the solutions (\ref{FGHR}) approach one of the
ground states in (\ref{Zground}).\footnote{More precisely, type of soliton
chooses boundary condition at the spatial infinity.
For example, the $(-1,0,0,0)$ soliton can exist in the first and the third
ground states in (\ref{Zground}) but not in the second.}
The topological charge in (\ref{topcur2}) depends on the winding number $n_i \; 
(i=1,2,3,4)$ and also which ground state soliton lives in, i.e. the behaviour
at the spatial infinity.

We have studied solitons for various values of $(\{ n_i \})$ and for the ``ground states"
(\ref{Zground}) by numerical calculation and verified that stable solutions
really exist.
Some of them are given in Figs.3, 4, 5 and 6.
Most of them have similar size with those in the NRCP$^1$ model.
Behaviours of $F(r)$, $G(r)$, etc, are different from each other and
it is very interesting to observe them by experiment of the bilayer quantum Hall state.
We have also observed that there are various (one-parameter family of) stable solutions
for fixed $(\{n_i\})$ and the boundary condition at the spatial infinity.
For example $(-1,0,0,0)$-type soliton, behaviours of $G(r)$ and $H(r)$
change whereas $F(r)$ and $R(r)$ are almost stable in various solutions.
Actually, the self-dual equations (\ref{equi}) have the following scale-invariant
component $U(r) \equiv u_1(r)-u_3(r)$ whose field equation is
\begin{equation}
{d^2U \over dr^2}+{1 \over r}{dU \over dr}+(n_1-n_2-n_3+n_4){\delta (r) \over r}=0.
\label{eqU}
\end{equation}
In real materials, Coulomb interactions between electrons fix this degrees 
of freedom.


\section{Conclusion}

\setcounter{equation}{0}

In this paper, we studied (pseudo-)spin textures in the QHE with internal symmetry.
Especially we examined relationship between the CSBE and the low-energy effective
field theory, the NRCP$^{N-1}$ models, and found that the NRCP$^{N-1}$
models describe quite well topological solitons in the original CSBE.
In the NRCP$^1$ model, the CP$^1$ variables couple with CS gauge field and 
there is the Hopf term, both of which give fractional spin and fractional statistics
to solitons.
Total spin and statistics of solitons coincide with those of solitons in the CSBE,
and numerical calculation shows that forms of these solitons are very
close to each other.

We also studied solitons in the CS gauge theory of the NRCP$^3$ model
and obtained topological soliton solutions.
Most of them have simliar size with those in the NRCP$^1$ model and 
are hopefully detectable by experiment.

QHS's which appear as a result of the condensation of these solitons are under study
and their properties will be reported in a future publication\cite{IOS}.


\end{document}